\documentclass[conference]{IEEEtran}
\IEEEoverridecommandlockouts

\usepackage{cite}
\usepackage{amsmath,amssymb,amsfonts}
\usepackage{algorithmic}
\usepackage{graphicx}
\usepackage{textcomp}
\usepackage{xcolor}
\usepackage{subfigure}
\def\BibTeX{{\rm B\kern-.05em{\sc i\kern-.025em b}\kern-.08em   T\kern-.1667em\lower.7ex\hbox{E}\kern-.125emX}}
\begin{document}

\title{Cascade Network Stability of Synchronized Traffic Load Balancing with Heterogeneous Energy Efficiency Policies\\

\thanks{The work is supported by EPSRC CHEDDAR: Communications Hub For Empowering Distributed ClouD Computing Applications And Research (EP/X040518/1) (EP/Y037421/1).}
}

\author{\IEEEauthorblockN{1\textsuperscript{st} Mengbang Zou}
\IEEEauthorblockA{\textit{School of Aerospace, Transport and Manufacturing} \\
\textit{Cranfield University, Bedford, UK}\\
m.zou@cranfield.ac.uk}
\and
\IEEEauthorblockN{2\textsuperscript{nd} Weisi Guo}
\IEEEauthorblockA{\textit{School of Aerospace, Transport and Manufacturing} \\
\textit{Cranfield University, Bedford, UK}\\
weisi.guo@cranfield.ac.uk}
\and
}

\maketitle

\begin{abstract}
Cascade stability of load balancing is critical for ensuring high efficiency service delivery and preventing undesirable handovers. In energy efficient networks that employ diverse sleep mode operations, handing over traffic to neighbouring cells' expanded coverage must be done with minimal side effects. Current research is largely concerned with designing distributed and centralized efficient load balancing policies that are locally stable. There is a major research gap in identifying large-scale cascade stability for networks with heterogeneous load balancing policies arising from diverse plug-and-play sleep mode policies in ORAN, which will cause heterogeneity in the network stability behaviour. 

Here, we investigate whether cells arbitrarily connected for load balancing and having an arbitrary number undergoing sleep mode can: (i) synchronize to a desirable load-balancing state, and (ii) maintain stability. For the first time, we establish the criterion for stability and prove its validity for any general load dynamics and random network topology. Whilst its general form allows all load balancing and sleep mode dynamics to be incorporated, we propose an ORAN architecture where the network service management and orchestration (SMO) must monitor new load balancing policies to ensure overall network cascade stability.
\end{abstract}


\section{Introduction}
Load balancing is an essential aspect of ensuring Quality-of-Service (QoS) and efficient usage of spectrum and power \cite{9873839, Shao24WLAN}. Reducing the operational expenditure (OPEX) and energy consumption of BSs means many BSs can reduce transmit power and handover users to neighbouring BSs, in what is called heterogeneous network cell zooming \cite{9130059}. Sleep mode operations in BSs further exploit the potential to improve efficiency. Even though numerous methods have been proposed to determine the amount of sharing load between BSs, the stability of load balance is usually modeled for small networks with homogeneous policies. Mixed policies in a heterogeneous network due to sleep mode and new ORAN plug-and-play policies as xApps \cite{ORAN2, 6GApp} is not considered.

\subsection{Literature in Load Balancing}
Current literature focus on designing standard compliant uniform policies to address load balancing challenges. This started with policies by engineering function design, where specific antenna tilt and azimuth is designed to compensate sleep mode patterns \cite{6502480}. More recent research focus on autonomous vehicles \cite{9896204}, and mobility load balancing (MLB) method addressed the problem by tuning logical BS boundaries autonomously \cite{alsuhli2021deep}. The traffic load of a BS is regularly monitored, and to alleviate the load, the BS shrinks its coverage area by modifying the parameter known as (CIO), which is assigned on a BS-pair basis \cite{park2017mobility}. Through the reduction of the coverage boundary of an overloaded BS, the edge users are handed over to an underloaded neighbour BS from the overloaded ones. Since the overloaded BS can hand over edge users to neighbour BSs by adjusting CIO values, numerous studies have framed the load balancing problem as an optimization problem and designed algorithms to optimize neighbour BS relational parameters CIO \cite{attiah2020load, kwan2010mobility, alsuhli2021mobility, wu2021load}. 

\subsection{Research Gap in Stable Load Balancing}
The stability of load balance in an ideal synchronized state means that all BSs maintain a high-efficiency load balance state even when under traffic perturbations. If the ideal load-balancing state is not stable, then a small perturbation will cause the constant handover of edge users between BSs, leading to poor QoS and the need for complex hysteresis methods \cite{8314691}. For an arbitrarily large network, instability can cause cascade handover (e.g., user handover in cell A causes unintended handovers in cell B, C, etc.) and hence energy efficiency problems, which has not been examined.

As such, the cascade stability of load balance is significant area of missing research to avoid unnecessary handovers. Our previous work \cite{moutsinas2019probabilistic} has shown an exact analytical relationship for global stability, relating local load balancing dynamics and network topology. However, it was based on the assumption that all BSs only have active mode and the load dynamics are homogeneous policy across the network.

As discussed earlier, the introduction of sleep mode, as well as other policies has been generally designed to realize substantial reduction of energy consumption \cite{wu2012traffic, wu2015base, renga2023trading}. The introduction of sleep mode causes stochastic heterogeneity in network handover behaviour and whether BSs can synchronize to the load balance state and maintain stability is elusive. Our previous work used regular patterns to examine cascade behaviour \cite{6502480} and later proposed an analysis framework based on Gershgorin circles of the eigenvalues \cite{moutsinas2019probabilistic}. However, these techniques are not valid in the case of heterogeneous policies. To bridge this gap, in this paper, we propose a framework to analyze the stability of the synchronized state of traffic load balancing in heterogeneous networks considering sleep mode with general load dynamic functions and random geographic networks.

\subsection{Novelty \& Contribution}
Here, we assume BSs may have a portfolio of load balancing dynamic policies, not all of which can ensure global network stability. The contributions in this paper are as follows:
\begin{enumerate}
    \item Establish a novel networked dynamic model for traffic load balancing considering both identical and non-identical load dynamics, which refers to the active mode and sleep mode of BSs (but generalisable later to any dynamic functions of policies). 
    \item Establish the cascade stability criteria of load-balancing process, relating to the load dynamics in each BS, coupling dynamics of load sharing between BSs and the network neighbour list topology. 
    \item Perform theoretical analysis on the eigen-spectrum of the Laplacian matrix of the load balancing network, which determines the convergence speed of BSs synchronizing to the ideal load-balancing state. We prove that the criterion is valid for any generalized load-balancing dynamics and random geographic networks. 
    \item Use stochastic geometry simulations to show examples of the proof and design the ORAN architecture to enable this monitoring processes in realistic networks.
\end{enumerate}

\begin{figure}[t]
    \centering
    \resizebox*{8cm}{!}{\includegraphics{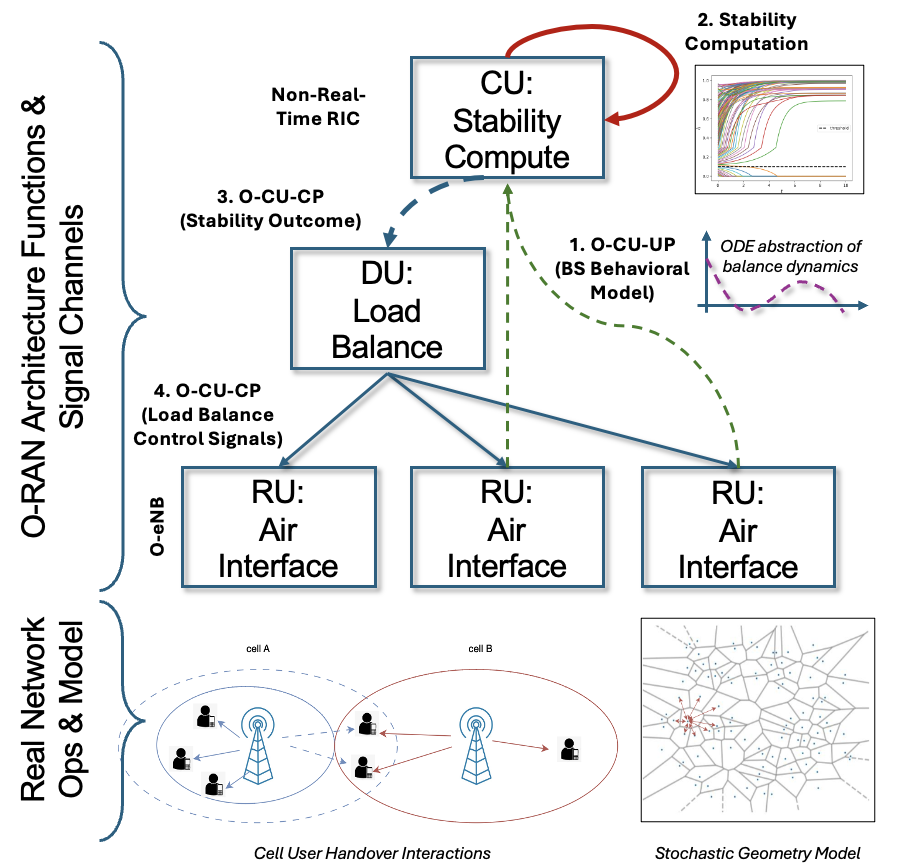}}
    \caption{(top) Open RAN architecture for stable load balancing: 1. BSs report their load balancing behavioral dynamics $f(.), g(.)$ to CU. 2. CU computes the cascade stability impact on whole network. 3. CU reports outcome to DU. 4. DU decides how to allow load balancing between RUs. (bottom) User-cell handover reality using equation~\eqref{equ: handover} and stochastic geometry model in paper.}
    \label{fig:oran}
\end{figure}

\section{System Setup}

\subsection{Architecture}
Architecturally, we refer to the Open RAN (ORAN) model to implement our network service management and orchestration (SMO) approach (shown in Fig.~\ref{fig:oran}top). In ORAN we expect innovators to design their own off-loading policies (inc. sleep mode) \cite{ORAN2} and whilst locally standard compliant and functionally desirable, it may cause global network cascade stability issues.

Here, we assume BSs may have a portfolio of load balancing dynamic policies, not all of which can ensure global network stability. As such, the steps are:
\begin{enumerate}
    \item BSs report their load balancing behavioral dynamics to Centralised Unit (CU). The assumption is that there exist some continuous mapping between policies and ODE function space.
    \item CU computes in non-real time the cascade stability impact on whole network using the framework in this paper. 
    \item CU reports stability outcome to Distributed Unit (DU), showing which BSs caused any stability issues. 
    \item DU decides how to allow load balancing between which Radio Unit (RU) (aka BS or cell). 
\end{enumerate}
We now present the model, assumptions, the cascade stability framework and results in stochastic geometry.

\begin{figure}[t]
    \centering
    \resizebox*{8cm}{!}{\includegraphics{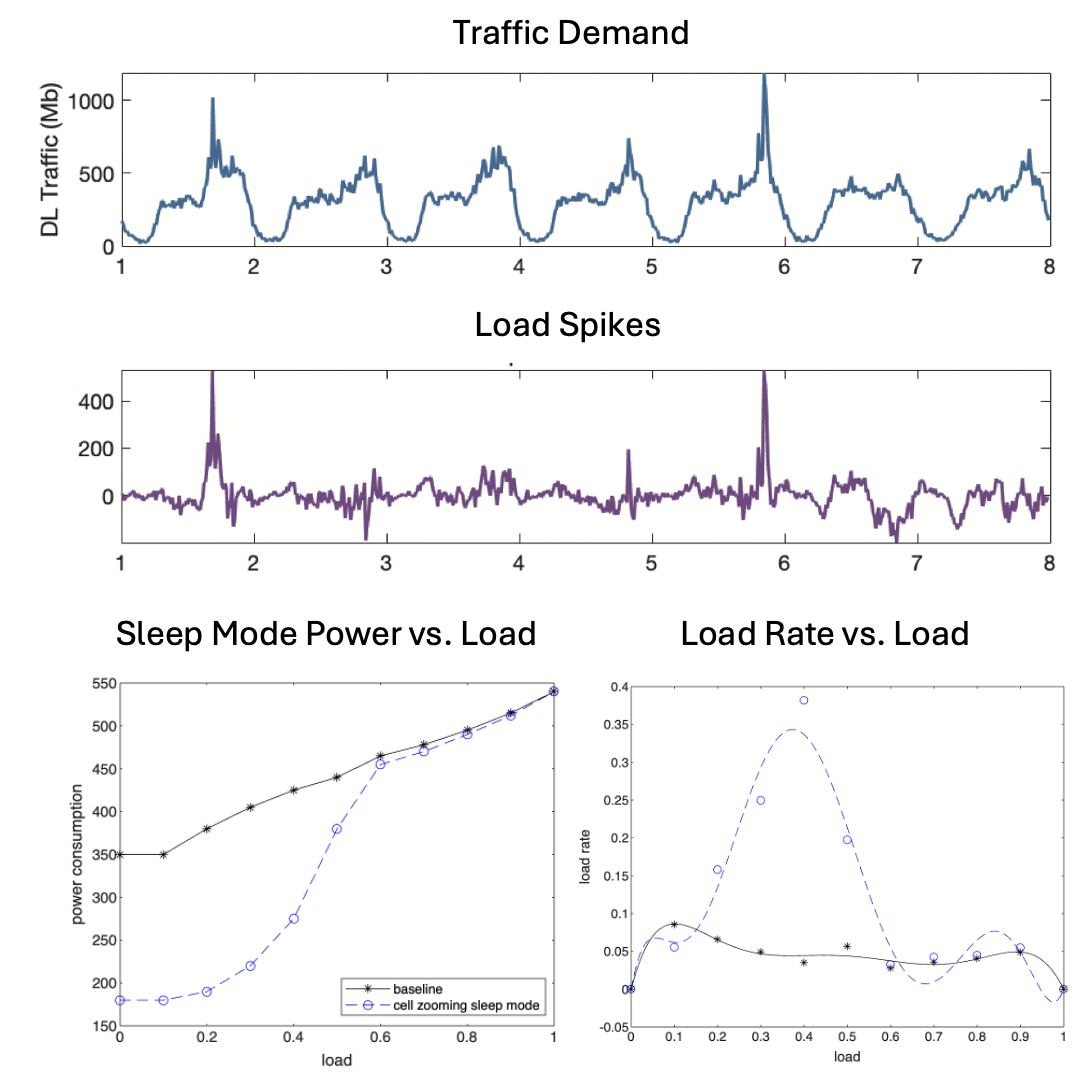}}
    \caption{(top) Real cellular traffic demand data and load spike data. (bottom-left) Base station performs a range of baseline power down, sleep mode, and dynamic cell zooming \cite{6502480}, trading off power consumption against traffic load. (bottom-right) Equivalent load rate of change against traffic load plot. The load dynamic plot data fitted with polynomial ordinary differential equation (ODE) to enable checking of cascade stability.}
    \label{fig:load_data}
\end{figure}

\subsection{Model \& Assumptions}
Consider the area covered by $N$ BSs. The load of a BS can be calculated by the utilization of physical resource blocks (PRBs) \cite{park2017mobility, kwan2010mobility} allocated to users $u$ from BS $i$ at time $t$ is
\begin{equation}
    B_{u,i}(t)=\frac{B_i(t)}{U_i(t)}\sum_{x=1}^{U_i(t)}(1/x),
\end{equation}
where $B_i(t)$ is the number of PRBs of BS $i$ can assign to users at time $t$, $U_i(t)$ is the number of users of BS $i$ at time $t$. The load of BS station $i$ at time $t$ can be calculated by 
\begin{equation}
    l_i(t) = \frac{\sum_{u=1}^{U_i(t)}\hat{B}_{u,i}}{B_i(t)}, 
\end{equation}
where $\hat{B}_{u,i}$ is the minimum number of PRBs that should be assigned to user $u$. $\hat{B}_{u,i}=Q_u/w(P_{u,i}, N_0)$, where $Q_u$ is the minimum data rate that needs to be provided, $w(P_{u,i}, N_0)$ is the rate function, $P_{u, i}$ is the power that user $u$ receive from BS $i$ and $N_0$ is the noise power. The data rate received by user $u$ is given by the rate function $w(\cdot)$ and the number of PRBs assigned to user $u$ according to \cite{park2017mobility}. Here, we do not concern ourselves with inter- or intra-cell handover differences, but are more concerned with the data-driven cascade stability theorems underpinning this area.


Without loss of generality, given load measurement data given above, the load dynamics in BS $i$ can be described by a continuous function
\begin{equation}
    \dot{l}_i=f_i(l_i),
\end{equation} which as shown in Fig.~\ref{fig:load_data} can be fitted to real data.

According to the 3GPP LTE specifications \cite{madelkhanova2022optimization}, a user initially served by BS $i$ will commence a handover request to some neighbour BS $j$ if the following condition holds
\begin{equation}\label{equ: handover}
    M_j+\theta_{j\to i}>Hys + M_i + \theta_{i \to j},
\end{equation}
where $M_i$ and $M_j$ are the measured values of reference signal received power (RSRP) from BS $i$ and $j$, $\theta_{i \to j}$ is the BS individual offset value of BS $i$ concerning BS $j$. The CIO is an offset that can be applied to alter the handover decision, which in effect changes the effective radius of BS $i$ concerning BS $j$. Here we set the offloading dynamics between two nodes as $g(l_i, l_j)$. $g(l_i, l_j)=0$ if $l_i = l_j$ and BSs $i, j$ are active. The overall network load balancing dynamics is the linear combinations of the local intra BS load dynamics $g(\cdot)$ and the inter BS load dynamics $f(\cdot)$. Referring to Fig.\ref{fig:oran}top, the aforementioned dynamics is reported from each RU to CU for cascade stability computation given below.

\begin{figure}[t]
    \centering
    \resizebox*{8cm}{!}{\includegraphics{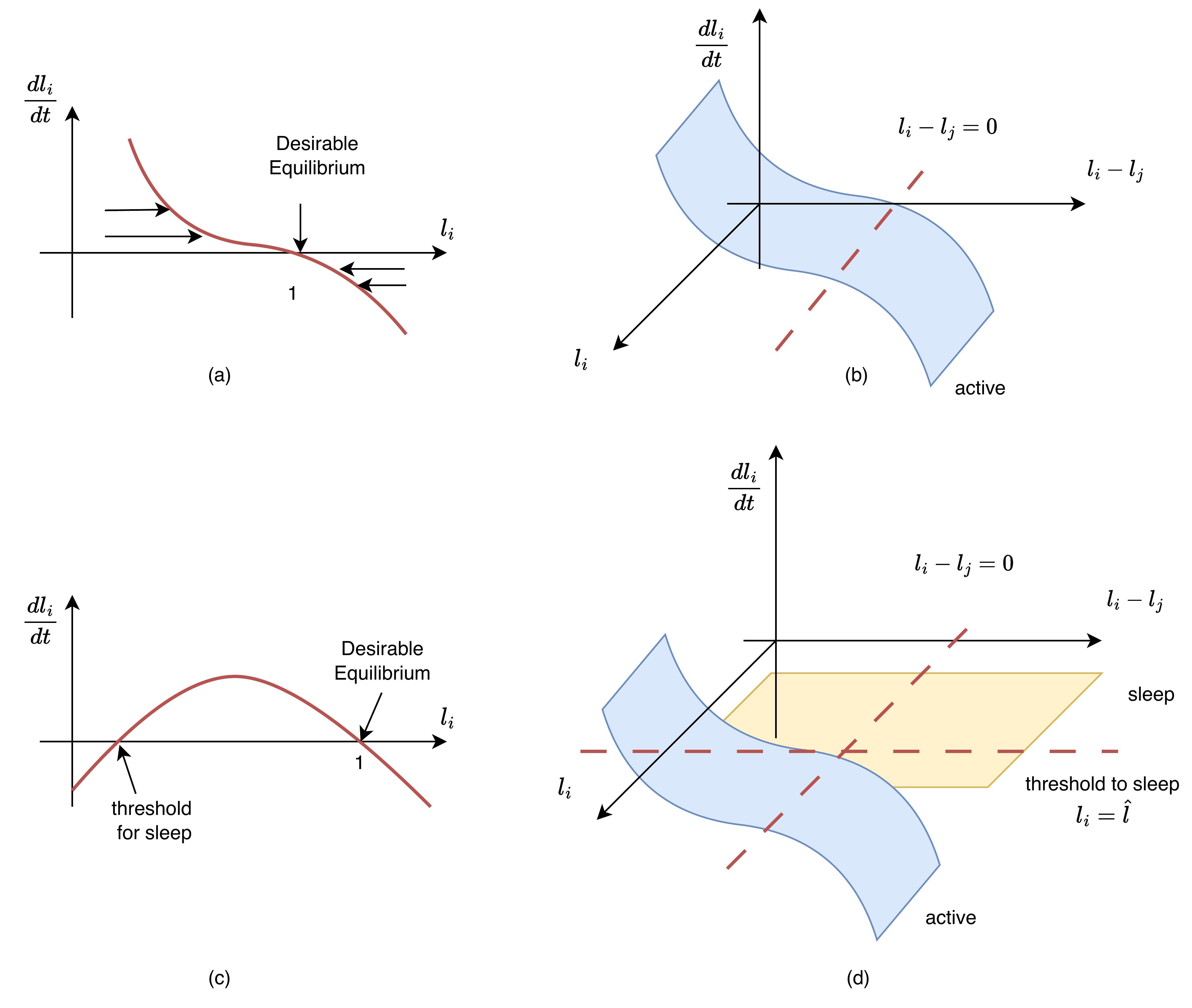}}
    \caption{Examples of potential dynamics: (a) The local load dynamics $l_i$ is a BS in active mode. If $l_i<1$, this underloaded BS should attract load to increase load. If $l_i>1$, then this overloaded BS should offload to other BSs. $l_i=1$ is a desirable equilibrium for maximum service efficiency. (b) The associated offloading dynamics $g(\cdot)$ of BS $i$ in active mode. (c) The local dynamics of BS $i$ in sleep mode. If $l_i$ smaller than the threshold of sleep, BS is sleeping and offloads to others. (d) is the associated offloading dynamics of BS in sleep mode. When the BS is sleeping, even $l_i<l_j$, BS $i$ still offload to other BSs. When $l_i$ is larger than the threshold, it is active and the offloading dynamics is similar to the BSs in active mode.}
    \label{fig: dynamics_modes}
\end{figure}

\subsection{Cascade Stability for Homogeneous Network}

The dynamics of each BS in active mode of the wireless network, which is regarded as a node in the complex network, is described by the equation
\begin{equation}\label{equ: coupling dynamics}
    \dot{l}_i = f(l_i)+\sum_{j=1,j \neq i}^Na_{ij}g(l_i, l_j),
\end{equation} 
where we assume that BSs in active mode have the same load dynamics. \textit{This is our strongest assumption: that there exist a continuous function that adequately describes load balancing algorithmic policies and acknowledge that how we realistically map policies to function space. We expect a data-fitting approach and this processes and its bifurcations are not tackled in this paper.}

The desirable equilibrium for maximum service efficiency is at $l_i=1, f(l_i=1)=0$ (fully loaded). The offloading dynamics $g(l_i, l_j)$ is decided by the term $l_i-l_j$ and $g(l_i, l_j)=0$ if $l_i=l_j$ (shown in Fig.~\ref{fig: dynamics_modes}). For any dynamics function $f(\cdot)$ and $g(\cdot)$, $l_i=1$ is always the equilibrium, which means that there exists a completely synchronized state for all nodes in this network i.e.,
\begin{equation}
l_1(t) = l_2(t) = \cdots = l_N(t) = s(t) = 1,
\end{equation}
where $s(t)$ is the solution of equation \eqref{equ: coupling dynamics}. A crucial problem is whether the synchronization manifold is stable in the presence of small perturbations $\delta l_i$. The linear evolution of small $\delta l_i$ can be obtained by setting $l_i(t)=s(t)+\delta l_i(t)$. $f(\cdot)$ and $g(\cdot)$ can be expanded by the first order Taylor series, i.e., $f(l_i)=f(s)+f'(s)\delta l_i$ and $g(l_i, l_j) = g(s,s) + g'_{l_i}(s,s)\delta l_i+g'_{l_j}(s,s)\delta l_j$. The linear variational equation for $\delta l_i$ is obtained by
\begin{equation}\label{equ: node coupling dynamics}
    \delta\dot{l}_i = f'(s)\delta l_i + \sum a_{ij} (g'_{l_i}(s,s) \delta l_i+g'_{l_j}(s,s) \delta l_j).
\end{equation}
Let $\textbf{l}=[l_1, l_2, \cdots, l_N]^{\top}$ and the equation~\eqref{equ: node coupling dynamics} can be written as 

\begin{equation}\label{equ: system coupling dynamics}
    \delta \dot{\textbf{l}} = f'(s)\delta\textbf{l}+g'_{l_i}(s,s)\textbf{D}\delta \textbf{l} + g'_{l_j}(s,s)\textbf{A}\delta\textbf{l},
\end{equation}
where $\textbf{A}$ is the adjacency matrix of the network and $\textbf{D}$ is the corresponding degree matrix. $a_{ij}=1$ if node $i$ connects to node $j$ and the diagonal elements $a_{ii}=0$, where $a_{ij}$ are the elements of matrix $\textbf{A}$. $\textbf{D}$ is a diagonal matrix where diagonal elements $d_{ii} = \sum_{i,j=1}^N a_{ij}$. Since $g(l_i, l_j)=0$ always has the solution $l_i=l_j$, $g(l_i, l_j)$ always contains the term $l_j-l_i$. So it is natural to assume that 

\begin{equation}
    g(l_i, l_j)=(l_j-l_i)(\sum p_nl_i^n+\sum q_ml_j^m+c).
\end{equation}
Then $g'_{l_i}(s, s)=-(\sum p_ns^n+\sum q_ms^m+c)$, $g'_{l_j}(s, s)=(\sum p_ns^n+\sum q_ms^m+c)$. Let set $g'_{l_i}(s, s)=-(\sum p_ns^n+\sum q_ms^m+c) = h(s, s)$. Then $g'_{l_j}(s, s)=(\sum p_ns^n+\sum q_ms^m+c)=-h(s, s)$. Then equation~\eqref{equ: system coupling dynamics} can be rewritten as 

\begin{equation} \label{equ: 7}
    \delta \dot{\textbf{l}} = f'(s)\delta\textbf{l}+h(s,s)\textbf{L}\delta \textbf{l},
\end{equation}
where $\textbf{L}=\textbf{D}-\textbf{A}$ is a Laplacian matrix corresponding to \textbf{A}. Laplacian matrix can be decomposed as $\textbf{L}=\textbf{S}^{-1}\Lambda\textbf{S}$, where $\Lambda$ is a diagonal matrix with diagonal elements $\lambda_1, \lambda_2, \cdots \lambda_N$ which are the eigenvalues of $\textbf{L}$. Further more, in Laplacian matrix $0 = \lambda_1 \leq \lambda_2 \leq \cdots \leq \lambda_N.$ Equation~\eqref{equ: 7} is equivalent to 
\begin{equation}
    \delta \dot{\textbf{l}} = f'(s)\delta\textbf{l}+h(s,s)\textbf{S}^{-1}\Lambda\textbf{S}\delta \textbf{l}.
\end{equation}
Left multiply $\delta\dot{\textbf{l}}$ with \textbf{S}, then
\begin{equation}
    \textbf{S}\delta \dot{\textbf{l}} = f'(s)\textbf{S}\delta\textbf{l}+h(s,s)\Lambda\textbf{S}\delta \textbf{l}.
\end{equation}
Let $\xi =[\xi_1, \xi_2, \cdots, \xi_N]^{\top}= \textbf{S}\delta\textbf{l} \in \mathbb{R}^N$, then
\begin{equation} \label{equ: decomposition}
    \dot{\xi} = f'(s)\xi+h(s,s)\Lambda\xi.
\end{equation}
By doing this, equation~\eqref{equ: decomposition} can be diagonalized into $N$ decoupled eigen-modes in the form 

\begin{equation}\label{equ: 11}
    \dot{\xi}_i = (f'(s)+h(s,s)\lambda_i)\xi_i.
\end{equation}

The evolution of $\xi_i$ with time is the solution of equation~\eqref{equ: 11}, $\xi_i = e^{(f'(s)+h(s,s)\lambda_i)t}\xi_0$. The system is stable to the equilibrium $s$ with $||\xi(t)||\to 0$, which requires $f'(s)+h(s,s)\lambda_i<0$. Since $\lambda_i \geq 0$, the stability is mainly decided by the value of $f'(s)$ and $h(s,s)$. 

Generally, in an active mode, $f'(s,s) < 0$ and $h(s,s) \leq 0$. $f'(s)+h(s,s)\lambda_i<0$. The synchronized state $s=1$ is stable. The synchronized speed is decided by the eigenvalues.

\begin{figure}[ht]
    \centering
    \subfigure[]{
    \includegraphics[width=0.23\textwidth]{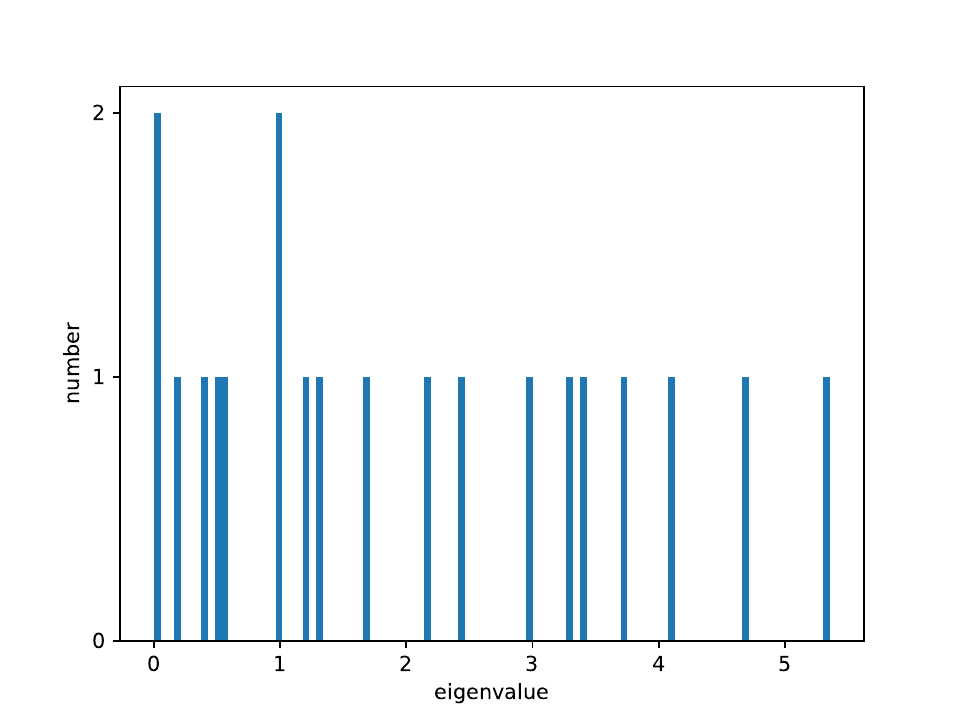}}
    \subfigure[]{
    \includegraphics[width=0.23\textwidth]{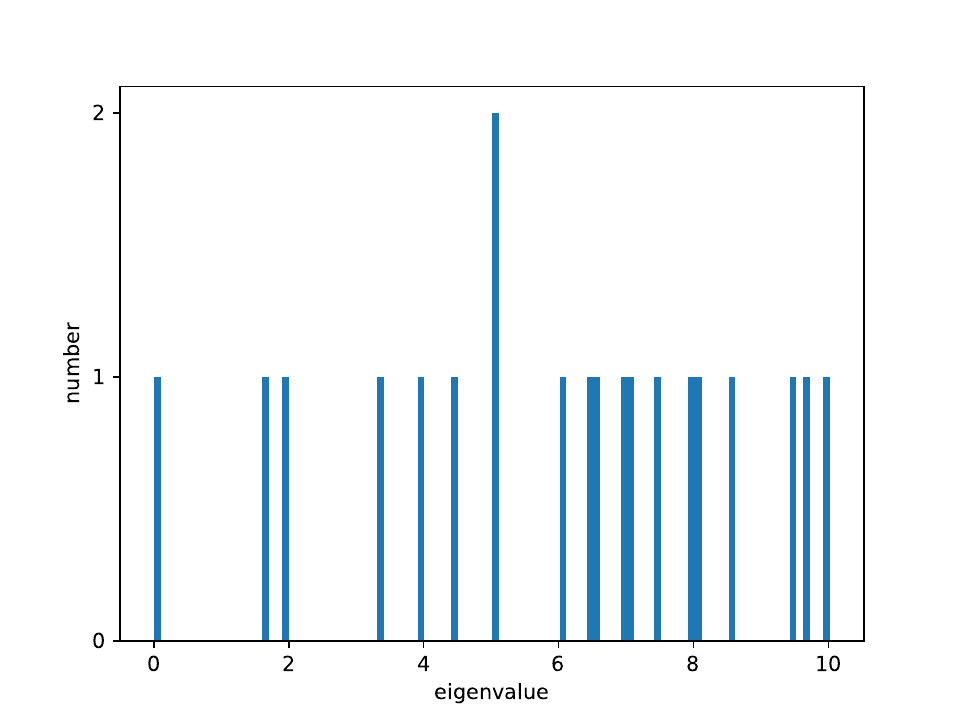}}

    \subfigure[]{
    \includegraphics[width=0.23\textwidth]{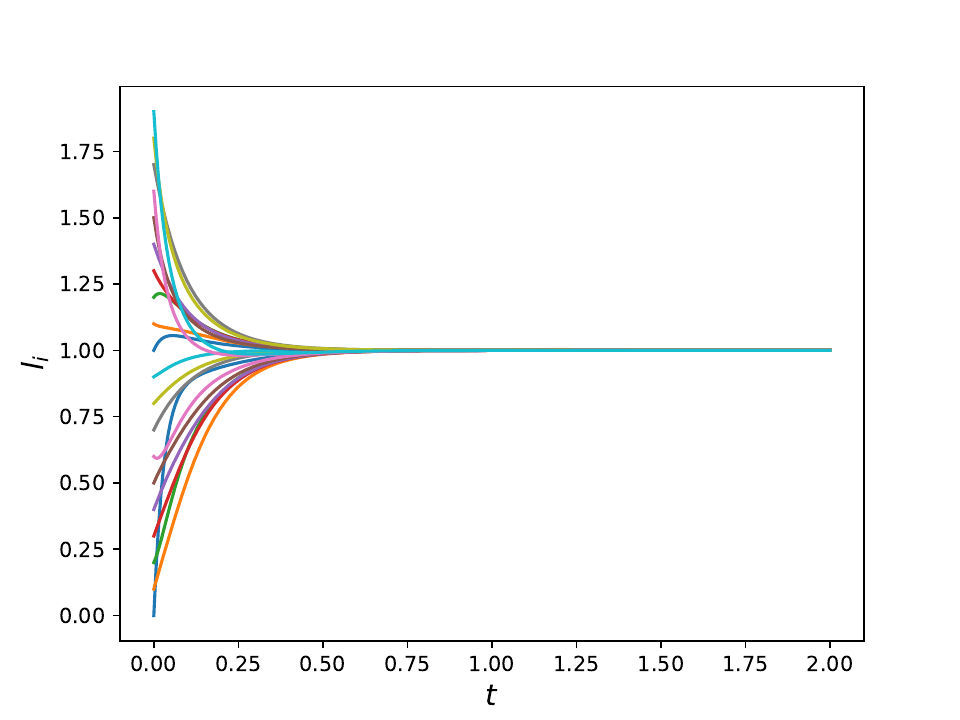}}
    \subfigure[]{
    \includegraphics[width=0.23\textwidth]{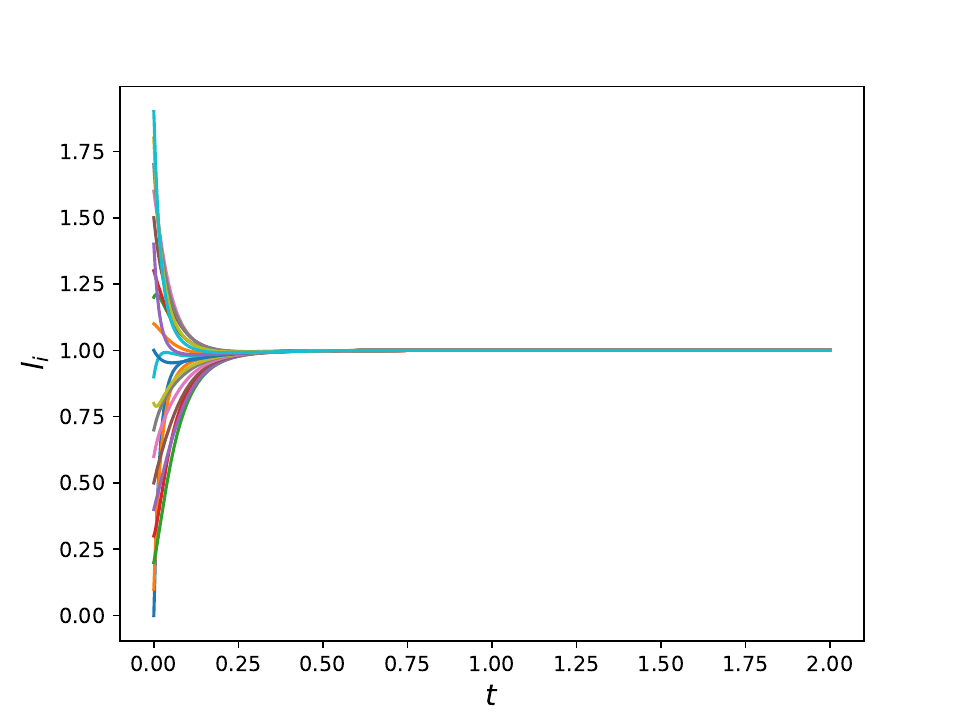}}
    \caption{(a) is the eigenvalue distribution of random network 1. (b) is the eigenvalue distribution of random network 2. (c) is the synchronization process of load balancing related to network 1. (d) is the synchronization process of load balancing related to network 2.}
    \label{fig: random_network1}
\end{figure} 

\begin{figure}[ht]
    \centering
    \subfigure[]{
\includegraphics[width=0.23\textwidth]{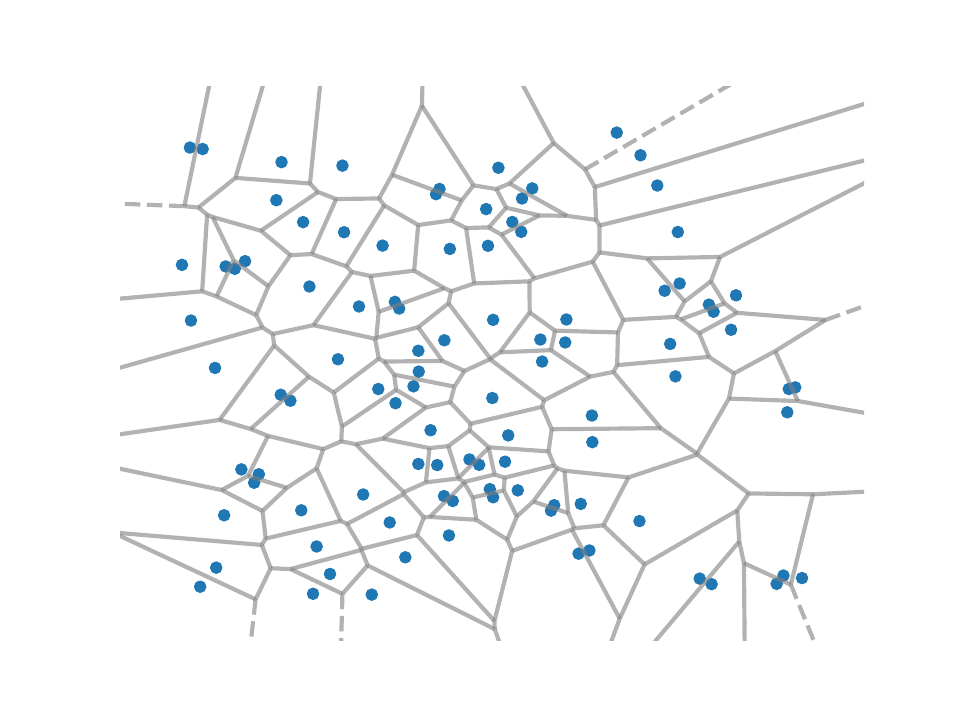}}
\subfigure[]{
\includegraphics[width=0.23\textwidth]{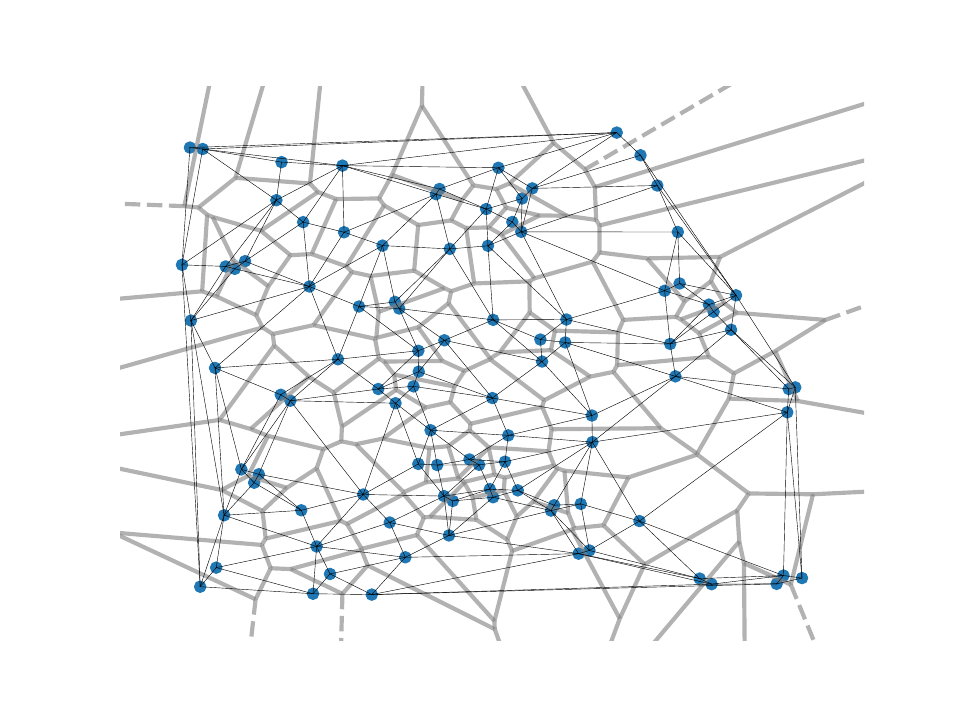}}
 
    \subfigure[]{
\includegraphics[width=0.23\textwidth]
{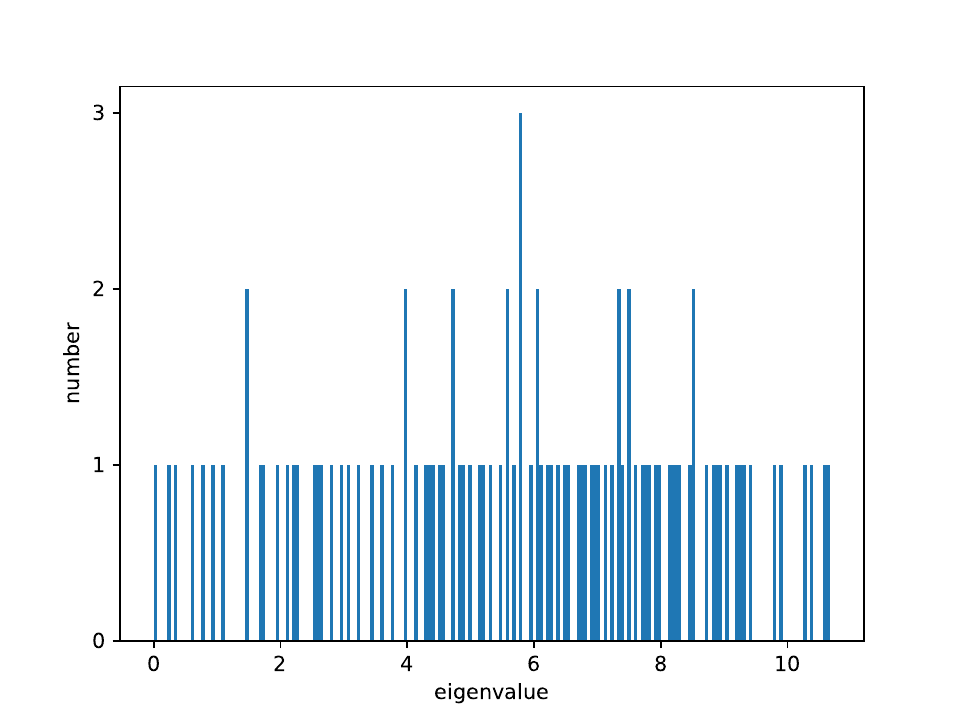}}
    \subfigure[]{
\includegraphics[width=0.23\textwidth]{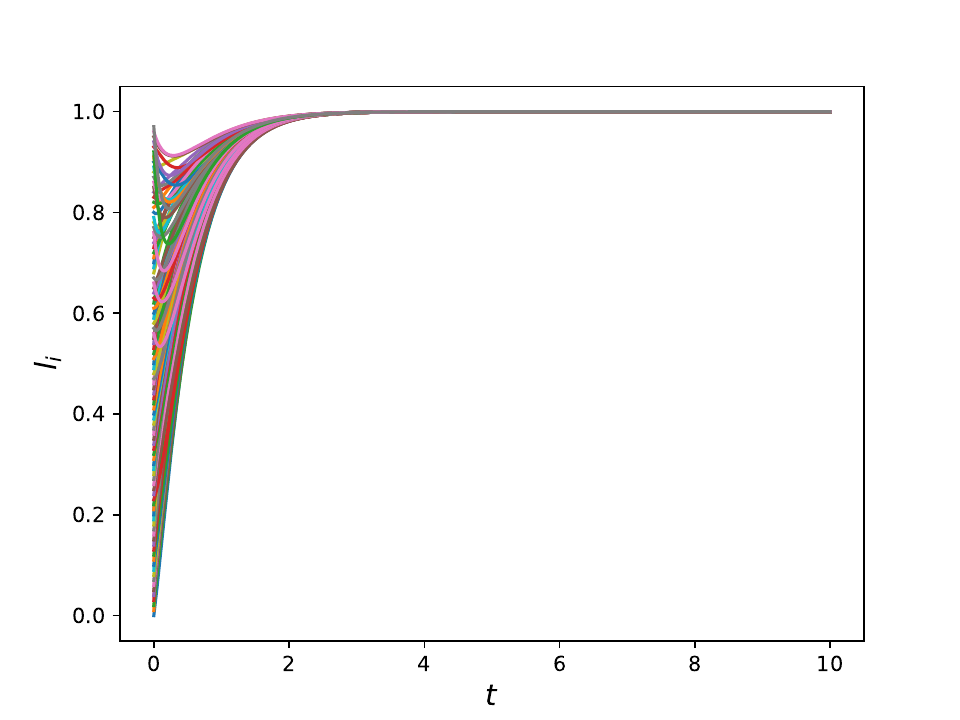}}
\caption{(a) The random geographic BSs generated by PPP, where blue nodes are BSs, grey lines are the boundary of BSs. (b) Complex network of BSs corresponding to (a), where black lines are offloading relations. (c) Eigenvalue distribution of the network. (d) Load balancing synchronization process.}
    \label{fig: PPP}
\end{figure}

\begin{figure*}[ht]
    \centering
    \subfigure[]{
\includegraphics[width=0.32\textwidth]{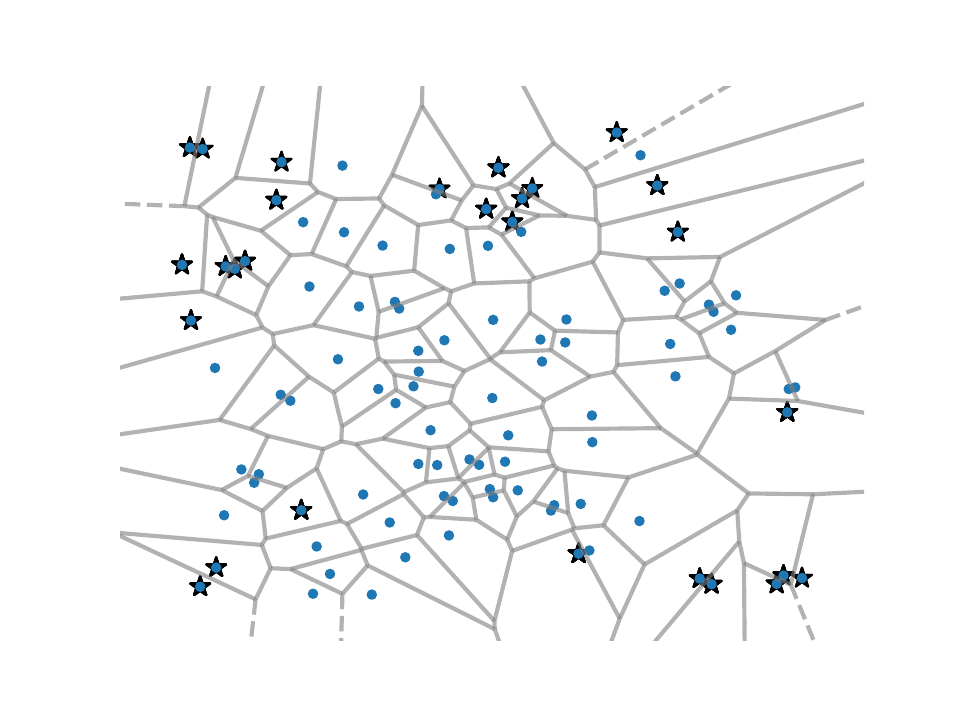}}
\subfigure[]{
\includegraphics[width=0.32\textwidth]
{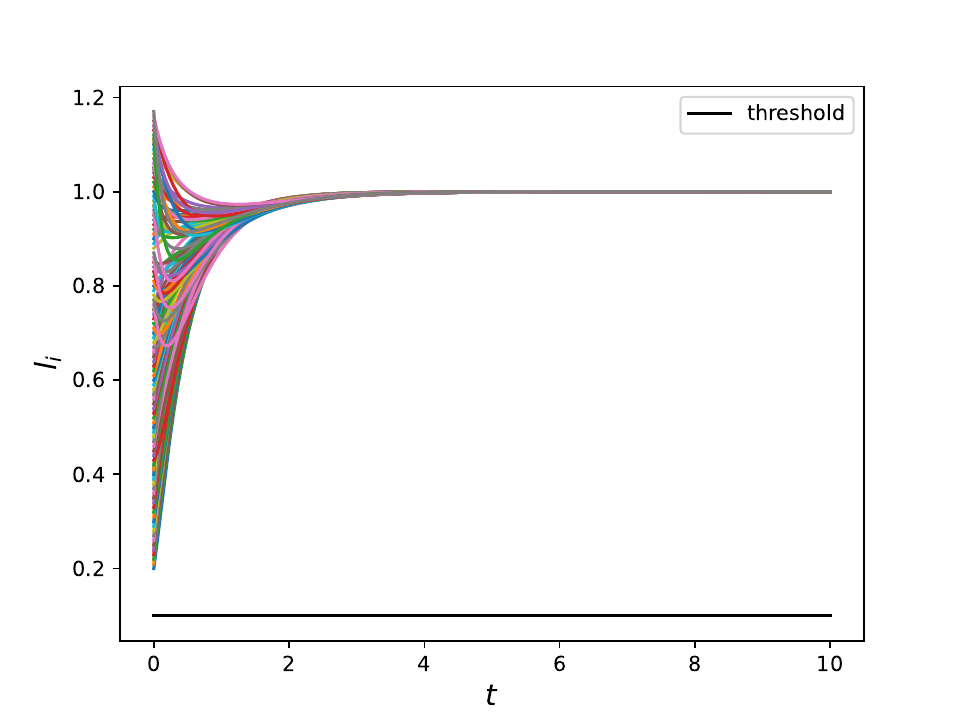}}
    \subfigure[]{
\includegraphics[width=0.32\textwidth]{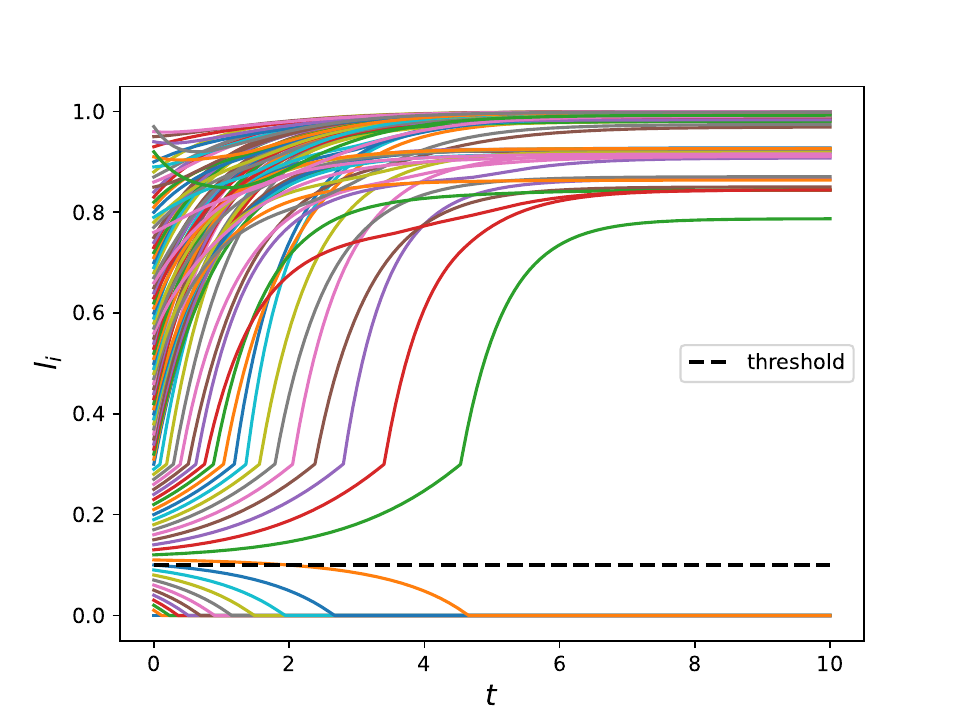}}
\caption{(a) The random geographic BSs generated by PPP, where blue circle nodes are BSs with active mode, star nodes are BSs with sleep mode. (b) load balancing process in a heterogeneous network. In this case, the threshold to switch off is small. (c) In this case, the threshold to switch off is larger than (b), and some BSs switch off at the beginning. BSs can not synchronize to the ideal load-balancing state $l_i=1$ and maintain stable.}
    \label{fig: hete}
\end{figure*} 

\subsection{Cascade Stability for Heterogeneous Network}
Now we consider the sleep mode (the local load dynamics and offloading dynamics shown in Fig.~\ref{fig: dynamics_modes}). Assume that $N_1$ BSs are in sleep mode and the remaining BS are in active mode. The coupling dynamics $g(l_i, l_j)$ have three different types: 1) dynamics between a BS in sleep mode and a BS in active mode; 2) dynamics between two BSs in active mode; 3) dynamics between two BSs in sleep mode. We use $g_{ij}(l_i, l_j)$ to represent different coupling dynamics between BS and assume that $g_{ij}(l_i, l_j)=g_{ji}(l_j, l_i)$, if $l_i, l_j > \gamma$, where $\gamma$ is the threshold value to switch. The load dynamics of BS with sleep mode is
\begin{equation}
    \left\{
             \begin{array}{lr}
             f_i(l_i) <0, \quad l_i<\gamma\\ 
             f_i(l_i) >0, \quad \gamma \leq l_i < 1  \\
             f_i(l_i) \geq 0, \quad l_i \geq 1\\
          
             \end{array}
\right.
\end{equation}
The offloading dynamics $g_{ij}(l_i,l_j)$ is 
\begin{equation}
    \left\{
             \begin{array}{lr}
             {g_{ij}(l_i, l_j)} <0, \quad l_i<\gamma\\ 
             g_{ij}(l_i, l_j) <0, \quad \gamma \leq l_i, \quad l_i>l_j  \\
             g_{ij}(l_i,l_j) \geq 0, \quad \gamma < l_i, \quad l_i \leq l_j\\
          
             \end{array}
\right.
\end{equation}

Then the dynamics of the system can be written as 
\begin{equation}
    \dot{l}_i= f_i(l_i) + \sum_{j=1,j \neq i}^Na_{ij}g_{ij}(l_i, l_j)
\end{equation}
Similar to equation~\eqref{equ: node coupling dynamics}, the deviations $\delta l_i=l_i(t)-s(t)$ satisfy
\begin{equation}\label{equ: 17}
    \delta\dot{l}_i = f_i'(s)\delta l_i + \sum a_{ij} (\frac{ \partial g_{ij} }{ \partial l_i }(s,s) \delta l_i+\frac{ \partial g_{ij} }{ \partial  l_j}(s,s) \delta l_j).
\end{equation}
Similar to equation~\eqref{equ: 7}, we have $\frac{ \partial g_{ij} }{ \partial l_i }(s,s) = -\frac{ \partial g_{ij} }{ \partial  l_j}(s,s)$. Let $\frac{ \partial g_{ij} }{ \partial l_i }(s,s)=h_{ij}(s,s)$. Then equation~\eqref{equ: 17} can be written as 

\begin{equation}
    \delta \dot{\textbf{l}} = \textbf{F} \delta\textbf{l}+ (\textbf{K}-\textbf{Q})\delta \textbf{l},
\end{equation}
where $\textbf{F}={\rm diag}\{f_1(l_1), f_2(l_2), \cdots, f_N(l_N)\}$, ${\rm diag} \{ \}$ represents the diagonal matrix. $\textbf{Q} = \textbf{A}\odot \textbf{H}$, where $\odot$ is a Hadamard product.
\begin{equation}
    \textbf{H} =  \begin{bmatrix}
    0 & h_{12}(s, s) & \cdots & h_{1N}(s, s) \\
    h_{21}(s, s) & 0  & \cdots & h_{2N}(s, s) \\
    \vdots & \vdots & \vdots & \vdots\\
    h_{N1}(s, s) & h_{N2}(s, s) &  \cdots & 0
    \end{bmatrix}
\end{equation}
$\textbf{K}$ is a diagonal matrix, where $k_{ii} = \sum_{j=1}^{N}q_{ij}$, $q_{ij}$ are the elements of $\textbf{Q}$.
According to Lyapunov's second method for stability, if there exists a Lyapunov function $V(\textbf{x})=\textbf{x}^{\top}\textbf{P}\textbf{x} > 0$ ($V(\textbf{x})=0$ only if $\textbf{x}=0$), where $\textbf{P}$ is positive definite, and $\dot{V}(\textbf{x})<0$, then the system is stable at the equilibrium. Let $\delta{\textbf{l}}=\textbf{x}$,then $\dot{V}(\textbf{x}) = \dot{\textbf{x}}\textbf{P}\textbf{x}+\textbf{x}^{\top}\textbf{P}\dot{\textbf{x}}$ can be expanded as 
\begin{equation}
  \dot{V}(\textbf{x})=(\textbf{x}^{\top}\textbf{F}^{\top}+\textbf{x}^{\top}(\textbf{K}^{\top}-\textbf{Q}^{\top}))\textbf{P}\textbf{x}+\textbf{x}^{\top}\textbf{P}(\textbf{F}\textbf{x}+(\textbf{K}-\textbf{Q})\textbf{x})
\end{equation}
Let $\textbf{P}=\textbf{I}$, where $\textbf{I}$ is identity matrix. Then
\begin{equation}
    \dot{V}(\textbf{x})=2(\textbf{x}^{\top}\textbf{F}\textbf{x}+\textbf{x}^{\top}(\textbf{K}-\textbf{Q})\textbf{x}).
\end{equation}

\begin{equation}
  \textbf{x}^{\top}(\textbf{K}-\textbf{Q})\textbf{x} = \sum_{i=1}^Nk_{ii}x_i^2 - \sum_{i, j=1}^N x_ix_jq_{ij}=\frac{1}{2}(\sum_{i,j=1}^Nq_{ij}(x_i-x_j)^2)  
\end{equation}

If $h_{ij}(s,s)<0$ and $f'_i(s) <0$, then $\textbf{x}^{\top}\textbf{F}\textbf{x}<0, \forall\textbf{x} \neq 0, \textbf{x}^{\top}(\textbf{K}-\textbf{Q})\textbf{x} \leq 0$. Therefore, $\dot{V}(\textbf{x})<0, \forall x\neq 0$. This indicates that $s=1$ is a stable equilibrium. 

If some BSs with lighter load tend to switch off,  then $g_{ij}(l_i, l_j) \ne g_{ji}(l_j, l_i)$, $\textbf{H}$ is not symmetric. Then $\textbf{K}-\textbf{Q}$ is not positive definite and $\dot{\textbf{V}}$ is not always negative. Therefore, the equilibrium $s=1$ is not stable. Only if BSs with sleep mode accumulate enough load to switch on, then the equilibrium $s=1$ is stable. Otherwise, the stability can not be guaranteed as well as the achievability of $s=1$.

Therefore, in a heterogeneous network where self-load dynamics and sharing dynamics are not identical, if $\frac{\partial g_{ij}}{\partial l_i}(s, s) <0$ and $f_i'(s)<0$, then the desirable equilibrium $l_i=1$ is stable and each BS can synchronize to this equilibrium at the end. However, if in some BSs the load is smaller than the threshold, $l_i< \gamma$, and these BSs switch off to sleep, the desirable equilibrium $l_i=1$ is unstable and BSs will converge to different states.

\section{Results}

First, we consider a nonlinear case, where all BSs are in active mode with the dynamics $\dot{l}_i = \alpha (1- l_i^2) + \beta\sum_{ij}a_{ij}(l_i-l_j)$, $\alpha > 0, \beta < 0$.  We generate small random networks, where nodes are BS sites and links are offloading relations, to verify our theory and show how eigen-spectrum affects the converge speed of load balance. In Fig. (\ref{fig: random_network1}), it shows that all BSs synchronize to the equilibrium $l_i=1$. The distribution of eigenvalues affects the convergence speed to equilibrium. Eigenvalues distribute near small values, leading to slow convergence speed. This simple case verified our theory that if $f'(s)<0$ and $g'_{l_i}(s,s)<0$, then the load balancing process will synchronize to the equilibrium $s$ and maintain stability.

We also generate large-scale random complex networks according to Poisson point process (PPP) \cite{saha20173gpp}
where nodes are BS sites and links are offloading relations. Nodes are connected with neighbour nodes with the probability $P$ to share loads. The results show that only if $f'(s)<0$ and $g'_{l_i}(s,s)<0$, the load balancing process will synchronize to the equilibrium and maintain stability.

Now, we consider a heterogeneous network with sleep mode. The local load dynamics of BSs in active mode is $f_1(l_i)=\alpha(1-l_i)$, where $l_i=1$ is the only equilibrium. Local load dynamics of BSs in sleep mode is $f_2(l_i)=\alpha((l_i-\gamma)(1-l_i))$, where $\gamma$ is the threshold of sleep mode. If $l_i<\gamma$, then the BS switches off and hands over users to other BSs even though this BS is underloaded. If BSs $i, j$ are active, $g(l_i, l_j)=\beta(l_i, l_j)$. BS $i$ is active and BS $j$ switches off, then $g(l_j, l_i)=\beta(l_j), g(l_i, l_j) = 0$. The results are shown in Fig~(\ref{fig: hete}). If the threshold $\gamma$ to switch off is small, all BSs are active and can synchronize to the ideal equilibrium $s=1$. Otherwise, the threshold $\gamma$ is large, some BSs switch off and sleep. BSs can not synchronize to the ideal equilibrium $s=1$. Some BSs converge to $s=0$ and some BSs converge to other states.

\section{Conclusion and Future Work}
In this paper,  we established the dynamic model for traffic load balance in wireless networks considering both identical and non-identical load dynamics, which refers to the active mode and sleep mode of BSs as well as other load balancing policies. In the framework of ORAN, we developed an architecture that allows the CU to check cascade stability of the whole network based on the policies RUs report to CU. We expect future innovators to design standard compliant policies, but they cannot be sure it won't cause wider network stability issues until the policy is reported to the ORAN CU for the checks designed in this paper.

The cascade stability criteria of load-balancing process is established in a transparent theoretical framework, mapping to the individual load dynamics in each BS, coupling dynamics of load sharing between BSs and the load balancing network topology. Our analysis indicated that for given load dynamics and offloading dynamics, the eigen-spectrum of the Laplacian matrix of the wireless network determines the convergence speed of BSs synchronizing to the ideal load-balancing state if all BSs are active. When some BSs switch off to sleep, BSs will converge to different states and the synchronized ideal load balancing state cannot be guaranteed. We verified our theory in generalized load-balancing dynamics and random geographic networks. 

In the future, we will explore how to design physics-informed neural networks to solve load-balancing problems based on the joint space between edge computing and wireless services \cite{9772936}, as well as experiment with current innovations such as BubbleRAN \cite{BubbleRAN}.

\bibliographystyle{IEEEtran}
\bibliography{ref}

\end{document}